\documentclass[english,aps,prl,show pacs,two column,superscript address,groupedaddress]{revtex4}
\usepackage[T1]{fontenc}
\usepackage[latin9]{inputenc}
\usepackage{graphicx,array,tabularx}
\usepackage{amstext}
\usepackage{hyperref}
\usepackage{bm,amssymb,amsmath,verbatim,cleveref}

\makeatletter

\@ifundefined{textcolor}{}
{%
 \definecolor{BLACK}{gray}{0}
 \definecolor{WHITE}{gray}{1}
 \definecolor{RED}{rgb}{1,0,0}
 \definecolor{GREEN}{rgb}{0,1,0}
 \definecolor{BLUE}{rgb}{0,0,1}
 \definecolor{CYAN}{cmyk}{1,0,0,0}
 \definecolor{MAGENTA}{cmyk}{0,1,0,0}
 \definecolor{YELLOW}{cmyk}{0,0,1,0}
}

\makeatother

\usepackage{babel}
\usepackage{color}

\begin{document}

\title{Efficient Estimation of Resonant Coupling between Quantum Systems}
\author{Markku P.V. Stenberg}
\email[]{markku.stenberg@iki.fi}
\affiliation{Theoretical Physics, Saarland University, 66123 Saarbr{\"u}cken, Germany}
\author{Yuval R. Sanders}
\affiliation{IQC and Dept. of Physics and Astronomy, University of Waterloo, 200 University Ave. W, Waterloo, ON, N2L 3G1, Canada}
\author{Frank K. Wilhelm}
\affiliation{Theoretical Physics, Saarland University, 66123 Saarbr{\"u}cken, Germany}
\affiliation{IQC and Dept. of Physics and Astronomy, University of Waterloo, 200
University Ave. W, Waterloo, ON, N2L 3G1, Canada}

\begin{abstract}
We present an efficient method for the characterization of two coupled discrete quantum systems, one of which
can be controlled and measured. For two systems with transition frequencies $\omega_q$, $\omega_r$, and
coupling strength $g$ we show how to obtain estimates of $g$ and $\omega_r$ whose error decreases exponentially in the number of measurement shots rather than as a power law expected in simple approaches. Our algorithm can thereby identify $g$ and $\omega_r$ simultaneously with high precision in a few hundred measurement shots. This is achieved by adapting measurement settings upon data as it is collected. We also introduce a method to eliminate erroneous estimates with small overhead. Our algorithm is robust against the presence of relaxation and typical noise. Our results are applicable to many candidate technologies for quantum computation, in particular, for the characterization of spurious two-level systems in superconducting qubits or stripline resonators. 
\end{abstract}
\pacs{03.65.Wj, 03.65.Yz, 03.67.-a, 03.67.Lx}
\maketitle

Parameter estimation in many microscopic and some macroscopic systems inevitably involves quantum measurements. This implies that parameters cannot be identified with a single measurement shot since the outcome of such a measurement is generally random. Instead, the standard approach is to determine ensemble
averages for many experiments and fit the parameters of certain quantitative models to those 
averages. The most common example for this is spectroscopy: it involves direct measurement of the energy splittings between quantum states in the form of resonances to incoming radiation. Typically, a large ensemble average is produced by gathering data from a large number of independent trials, either simultaneously on an ensemble of molecules (in nuclear magnetic resonance~\cite{levitt08}) or from many repetitions of a specific experiment (in optical spectroscopy of single molecules, quantum dots, or superconducting qubits~\cite{demtroder08}).

While being reliable in many contexts, this approach is often too resource intensive. Specifically, the error in the estimate of a single expectation value at a fixed measurement setting decreases in proportion to $M_{r}^{-\frac{1}{2}}$ after $M_r$ measurement shots. Moreover, many choices of measurement settings are usually required for 
complex measurement tasks. Such slowness of parameter estimation can also turn into imprecision 
in the estimate if the parameters of interest drift as a function of time, broadening spectroscopic signatures.

Imprecision in system characterization is particularly problematic for quantum information processing applications. These require extremely precise logic operations, usually implemented as pulses. The pulse parameters such as length, amplitude, and carrier frequency, depend on the system parameters. In manufactured solid state qubits, this is rather central as they are subject to fabrication uncertainty.

\begin{figure}
\includegraphics[width=0.5\textwidth]{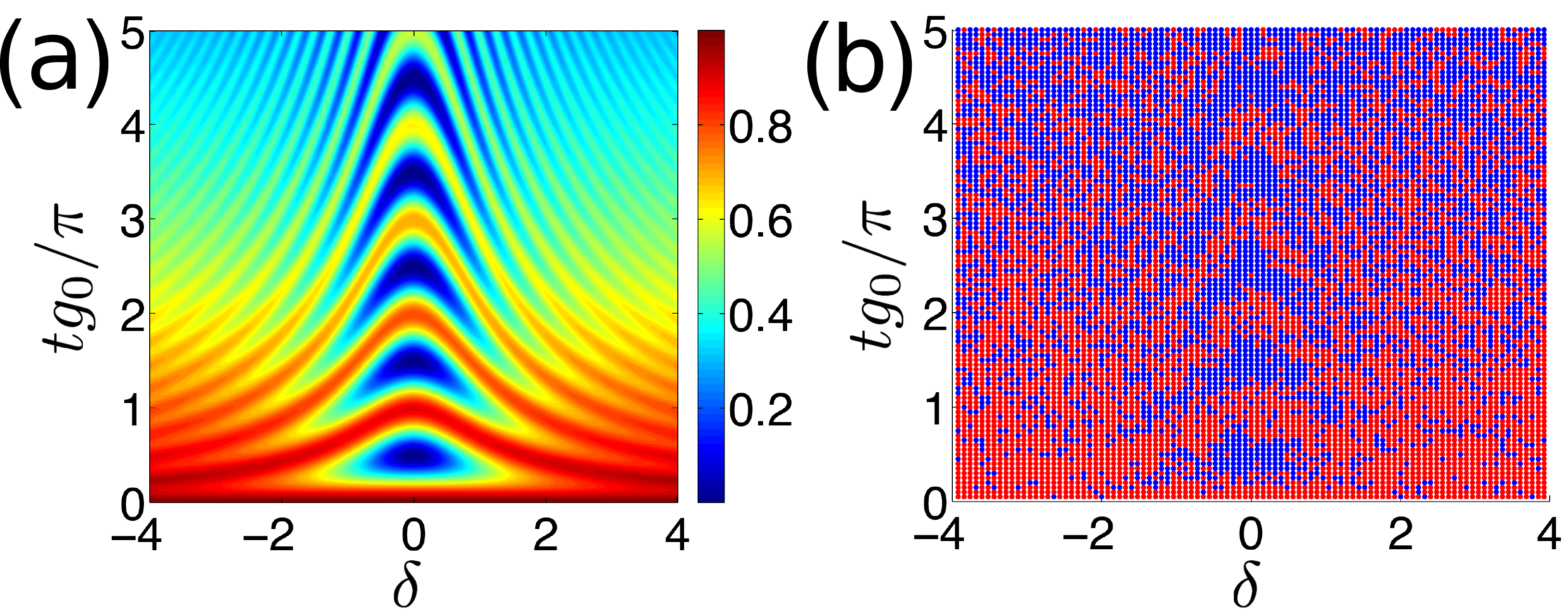}
\caption{(color online) (a) Theoretically obtained swap spectrum in frequency--waiting time plane. Here, \mbox{$\delta=(\omega_{\rm q}-\omega_{r,0})/({2g_0})$}, with $\omega_q$ the qubit frequency, $\omega_{r,0}$ the resonator frequency, and $g_0$ the coupling strength. The color scale of the swap spectrum represents the probability of the qubit being in its excited state. (b) Outcomes of a set of simulated single-shot measurements: blue the ground state; red the excited state.}
\label{fig:fig1}
\end{figure}

In this Letter, we demonstrate that advanced spectroscopy can be performed far more efficiently. Our results are particularly relevant for the ``tune-up'' stage in quantum information processing, {\it i.e.}, for the initial calibration of the system and maintaining it. We use a controlled qubit to characterize the uncertain frequency of another mode $\omega_r$ that is coupled to the qubit with uncertain coupling strength $g$. Our algorithm employs modern Bayesian inference techniques to choose informative experimental settings while remaining computationally feasible. We demonstrate that our approach is robust against experimental imperfections \cite{granade12,wiebe14b}. 

Our work belongs to the context of estimating an {\it a priori} unknown Hamiltonian \cite{sergeevich11,granade12,ferrie13,wiebe14a,wiebe14b}. Modern inference techniques have also been developed, e.g., for phase estimation \cite{giovannetti04,higgins07,berry09,hentschel10,xiang11,hentschel11,joo11,yonezawa12,waldherr12,nusran12} and state estimation \cite{huszar12,okamoto12,kravtsov13}. These techniques are able to use each bit of data obtained from experiment, instead of inferred expectation values only. This is both faster and more precise than the conventional approach involving ensemble averages.

Bayesian inference has been applied to identify qubit Hamiltonians in a standard tomographic setup using fixed, evenly spaced waiting times $t$ between preparation and measurement \cite{schirmer09}. In extracting an unknown frequency of a qubit, significant advantage can be achieved through an adaptive algorithm \cite{sergeevich11,ferrie13} that updates the measurement setting during the experimental data collection and tends to choose exponentially increasing, rather than evenly spaced, waiting times. Such a problem is equivalent to extracting $g$ in a model discussed below when the frequency $\omega_r$ is known. The assumption that $\omega_r$ is known, however, limits the applicability of the model. The situation where both $g$ and $\omega_r$ are initially unknown is more widely and practically applicable. To solve the latter problem, we deliver a strategy that chooses both $t$ and the qubit frequency $\omega_q$ to achieve near-optimal scaling of errors in the estimates.

A qubit coupled to a resonator is described by the Jaynes-Cummings Hamiltonian
\begin{equation}
\hat{H}_{{\rm JC}}=\frac{\hbar\omega_{q}}{2}\hat{\sigma}_{z}+\hbar\omega_{r}\left(\hat{a}^{\dagger}\hat{a}+\frac{1}{2}\right)+\hbar g\left(\hat{\sigma}_{+}\hat{a}+\hat{\sigma}_{-}\hat{a}^{\dagger}\right),
\end{equation}
where $g\ll \omega_r$. This model is broadly applicable to potential quantum computing technologies \cite{h06b09i99} 
and reproduces also the dynamics of spurious two-level systems\cite{neeley08,palomaki10,shalibo10}, a notorious source of decoherence for superconducting qubits.

The standard method to estimate $g$ and $\omega_{r}$ is called swap spectroscopy \cite{mariantoni11}.
To understand swap spectroscopy, it helps to picture a single measurement setting as a single point in a figure such as Fig.~\ref{fig:fig1}(a) or \ref{fig:fig1}(b). One starts by preparing the qubit in the excited state and the cavity in its ground state (typically by moving $\omega_{q}$ far away from $\omega_{r}$ and then exciting it with an external pulse). Then $\omega_{q}$ is 
fixed to a chosen value that in Fig.~\ref{fig:fig1} determines the horizontal coordinate (see the caption).
The system is allowed to evolve a time $t$ (vertical coordinate of Fig.~\ref{fig:fig1}) after which the qubit is measured in the $\hat{\sigma}_{z}$ basis. The system is then reset to its ground state before the next measurement.

As the Jaynes-Cummings Hamiltonian conserves the total number of excitations $\hat{N}=\hat{a}^{\dagger}\hat{a}+\frac{1}{2}\left(\hat{\sigma}_{z}+\hat{1}\right)$ we can describe the excitation oscillating between the qubit and the resonator in the single-excitation subspace by the Hamiltonian $\hat{H}'=\frac{\Delta\omega}{2}\hat{\eta}_{z}+g\hat{\eta}_{x}$, where $\hat{\eta}$ are Pauli matrices and the detuning $\Delta\omega=\omega_{q}-\omega_{r}$.
This form of the relevant Hamiltonian occurs, in addition to the applications mentioned above, also in magnetic resonance spectroscopy. For the Hamiltonian ${\hat H}'$ the probability of the qubit being in its excited state is
\begin{eqnarray}
&P_{\omega_q,t}\left(1|g,\omega_{r}\right)=\frac{1}{2}\left(\frac{4g^{2}}{\omega_{R}^{2}}\cos\omega_{R}t+1+\frac{\Delta\omega^{2}}{\omega_{R}^{2}}\right),
\label{eq:ex_prob}
\end{eqnarray}
with $\omega_{R}=\sqrt{\Delta\omega^{2}+4g^{2}}$. The Supplemental Material generalizes this formula to account for qubit relaxation.

In conventional swap spectroscopy, the measurement is repeated at a setting $\left(\omega_{q},t\right)$ in order to establish an ensemble to calculate the relative frequency of the excited state and, hence, approximate its probability. 
To estimate $g$, usual swap spectroscopy first seeks $\omega_q$, where $\Delta\omega = 0$, and where 
the Chevron pattern of the oscillating excitation probability has a maximum visibility; cf. Fig.~\ref{fig:fig1}. Measuring 
the angular frequency of these oscillations then yields $2g$ by Eq.~\eqref{eq:ex_prob}.
\begin{figure}
\includegraphics[width=0.45\textwidth]{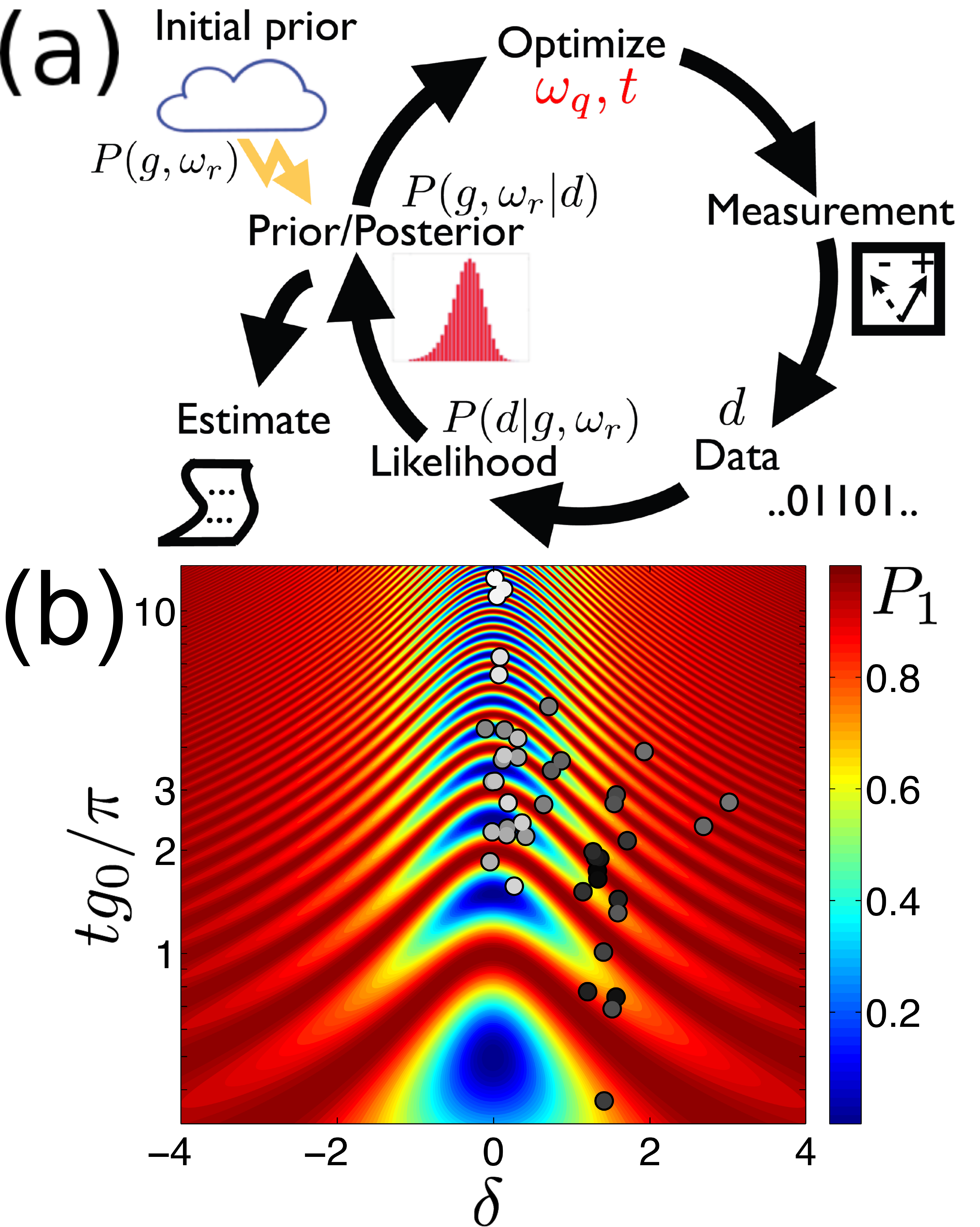}
\caption{(color online) (a) Illustration of an adaptive Bayesian inference scheme. (b) Exemplary trace of adaptive measurements overlaid on swap spectrum in the frequency--waiting time plane. The order of the measurement shots is denoted by their color (from black to white). Here $\delta=(\omega_{\rm q}-\omega_{r,0})/(2g_0)$, with $\omega_q$ the qubit frequency, $\omega_{r,0}$ the resonator frequency, and $g_0$ the coupling strength. The color scale for the swap spectrum represents the probability for the qubit to be in its excited state. Note the logarithmic scale on the time axis.}
\label{fig:fig2}
\end{figure}

Our algorithm, rather than establishing ensemble averages, chooses a new pair $(\omega_q,t)$ after each
measurement step. We employ Bayes' theorem, which reads 
\begin{equation}
P(g,\omega_{r}|d)=\frac{P(d|g,\omega_{r})P(g,\omega_{r})}{P(d)},\quad d=0,1
\label{eq:bayes_theorem}
\end{equation}
for our system. This formula can be understood as a rule for iterative learning of the
parameters. One starts with an initial probability distribution, or prior, $P(g,\omega_{r})$ that describes one's {\it a priori} conception about the uncertain parameters. Based on the measurement outcome $d$, cf.~Fig.~\ref{fig:fig2}(a), one can use Bayes' theorem to update the probability distribution into a posterior based on the likelihood of the data $P(d|g,\omega_{r})$ according to conjectured model parameters $g$ and $\omega_{r}$. The normalization factor in the denominator can be calculated via the integral $P(d)=\int P(d|g,\omega_{r}) P(g,\omega_{r}) dg d\omega_r$. The estimate is obtained from the mean value of the posterior. The posterior is then identified as the prior for the next measurement. Thus each measurement outcome is immediately incorporated into our knowledge of the system.

\begin{figure}
\includegraphics[width=0.4\textwidth]{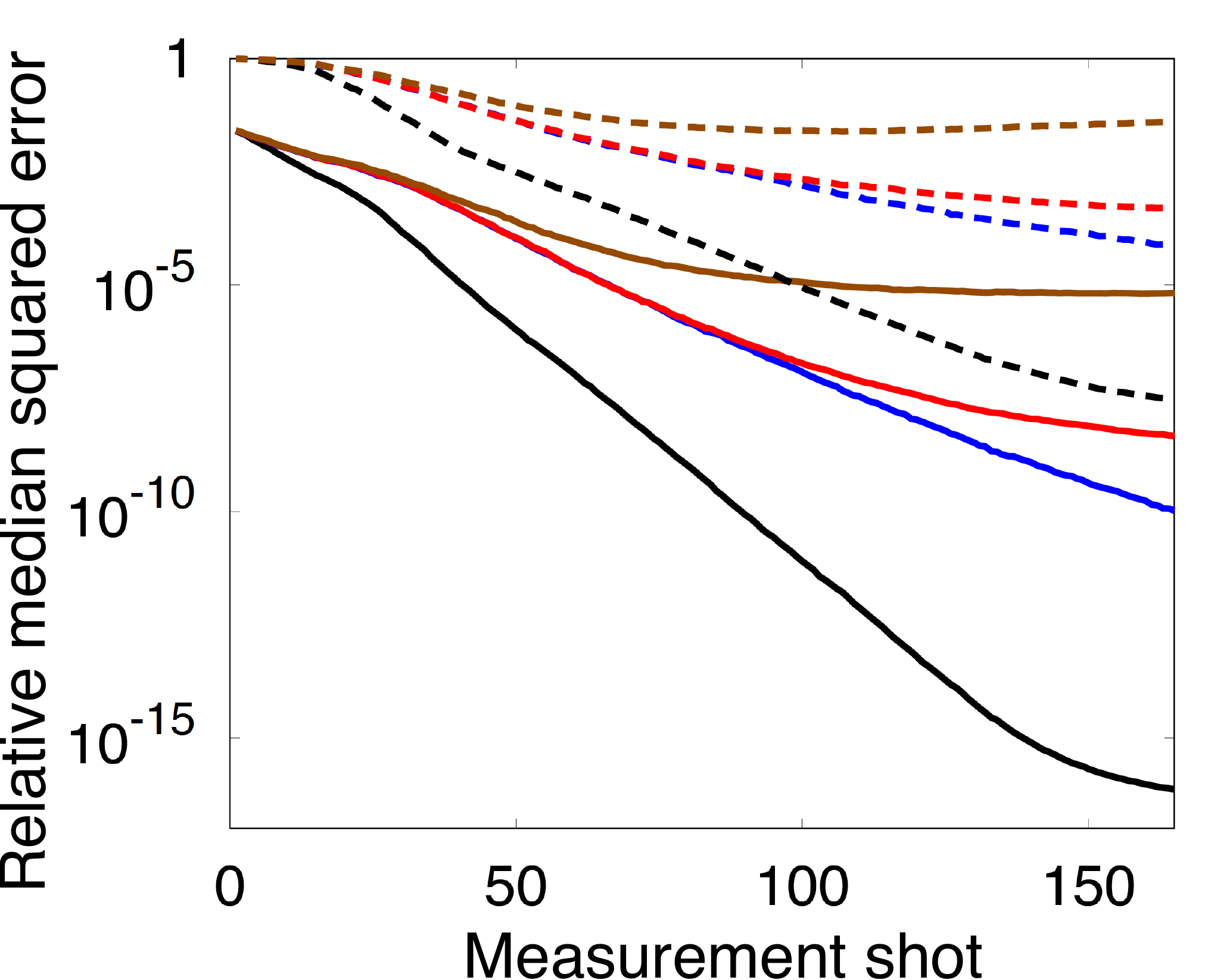}
\caption{(color online) Relative median squared error of the estimate of $g$ (solid) and $\omega_r$ (dashed)
calculated from an ensemble of 10 000 simulated samples (see text). The curves correspond to $T_1=\infty$ and $P_e=0$ (black), $T_1=\infty$ and $P_e=0.1$ (blue), $T_1\mu_g=2000\pi$ and $P_e=0.1$ (red), and $T_1\mu_g=40\pi$ and $P_e=0.1$ (brown). Here, $T_1$ is the relaxation time, $P_e$ is the probability for a readout error, and $\mu_g$ the mean of $g$ over the ensemble.}
\label{fig:fig3}
\end{figure}

\begin{figure}
\includegraphics[width=0.5\textwidth]{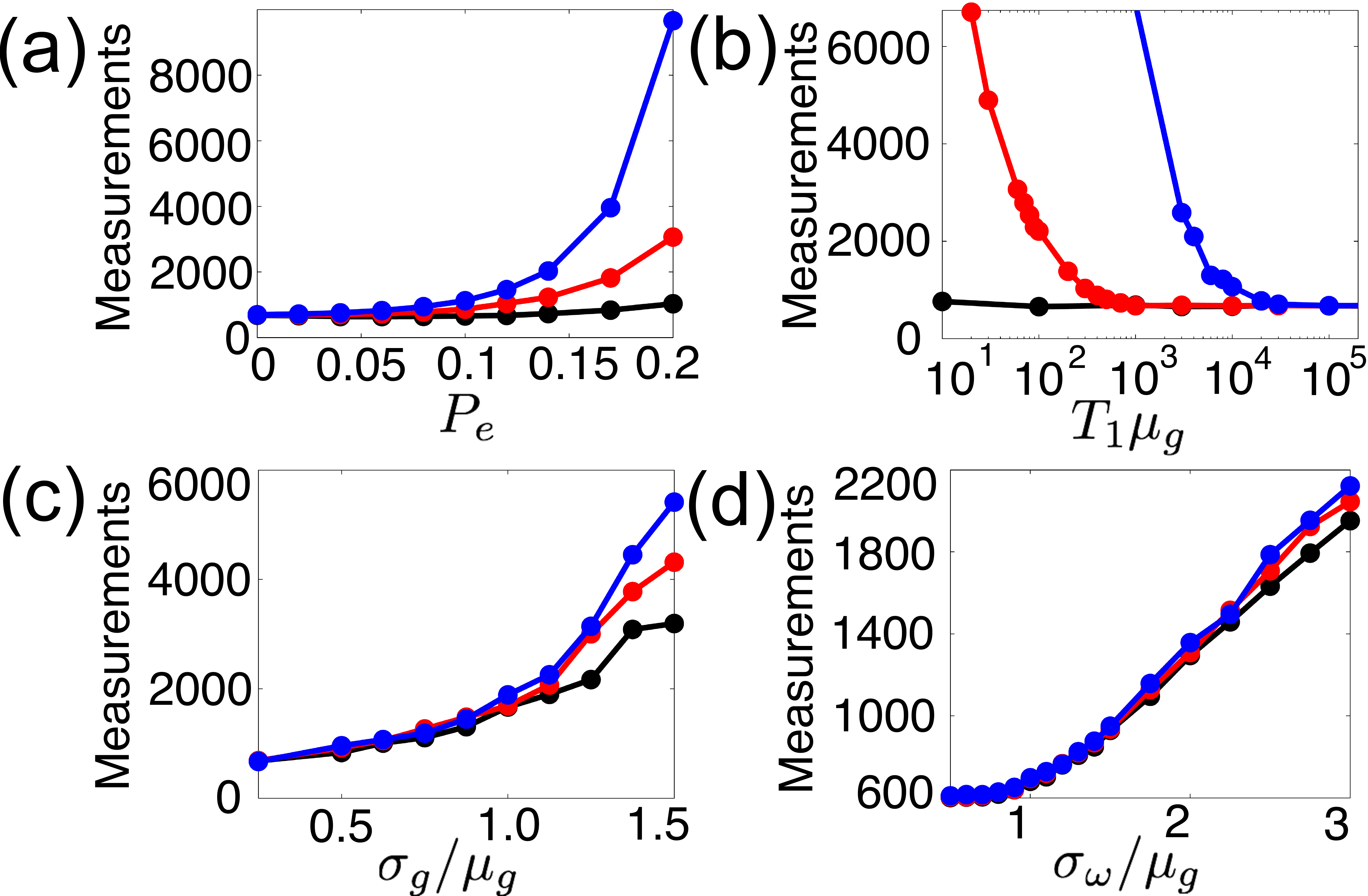}
\caption{(color online) The average number of measurement shots required to meet the desired level of relative 
mean squared error $\mathcal{E}^2_g$: $10^{-4}$ (black), $10^{-7}$ (red), and $10^{-10}$ (blue). The number of measurements presented  as a function of  (a) probability of readout error $P_e$, (b) relaxation time $T_1$, (c) standard deviation $\sigma_g$ over the initial prior, and (d) standard deviation $\sigma_\omega$ over the initial prior.}
\label{fig:fig4}
\end{figure}

We now want to optimize the measurement settings based on the current knowledge about $\omega_r$ and 
$g$. In principle, for each adaptation step one can maximize utility (e.g., a negative trace of the
posterior covariance matrix or information gain $I$ \cite{sivia06}) of the next shot as if it were the last 
in the series, a so-called ``greedy'' algorithm. Computationally however, maximizing utility between the shots is quite unwieldy. Moreover, to optimize the whole series of measurement shots, {\it i.e.}, to optimize globally, it is not 
sufficient to optimize greedily, {\it i.e.}, locally. We have therefore studied information gain $I(\omega_q,t)$ as a function of the control parameters with different parameter values $(g,\omega_r)$ and different priors. These considerations (which we carried out off-line rather than in real time parallel with parameter estimation) suggest a strategy in which $t\sim 1/\sigma_g$ and $\omega_q - \mu_\omega \sim \sigma_\omega$. With respect to the prior, $\mu_{\omega}$ is the mean of $\omega_r$ and $\sigma_g$ ($\sigma_\omega$) is the standard deviation of $g$ ($\omega_r$). Our measurement strategy chooses the $M$th measurement setting $(\omega_q,t)$ according to the rule
\begin{eqnarray}
t&=&\begin{cases}
\frac{ar_1}{\sigma_{g}}\quad &{\rm if}\ M \le M_0\\
\frac{|a+bz|}{\sigma_{g}}\quad &{\rm if}\  M > M_0
\end{cases},\nonumber\\
\omega_q&=&\begin{cases}
\mu_\omega+\left(r_2-\frac{1}{2}\right)\mu_g\quad &{\rm if}\ M \le M_0\\
\mu_\omega+c\left(r_2-\frac{1}{2}\right)\sigma_{\omega}\quad &{\rm if} \ M > M_0.
\end{cases}
\label{eq:update_rule}
\end{eqnarray}
Here, $a=1.57$, $b=0.518$, and $c=3.0$ are numerical constants that we have found to yield a robust and 
efficient strategy. Furthermore, $z$ is a standard normal deviate and $r_{1,2}$  are uniform random variables on the interval $[0,1]$. This strategy is the central result of our Letter. The measurement settings with $M\le M_0$ are chosen more uniformly to obtain a unimodal posterior in the beginning of the series \cite{ferrie13}, which makes the strategy more effective. We choose $M_0=15$ which we found to be sufficient especially for our prior and the parameters considered in Figs.~\ref{fig:fig3}, \ref{fig:fig4}(a), and \ref{fig:fig4}(b). Figure \ref{fig:fig2}(b) illustrates an exemplary trace of measurement shots chosen adaptively according to Eq.~(\ref{eq:update_rule}). Even though the
algorithm adaptively chooses measurements based on the current uncertainty in the Hamiltonian, it is different from ``particle guess heuristic'' (PGH) of Refs.~\cite{wiebe14a,wiebe14b} since it does not attempt to counteract the time evolution $e^{-i\hat{H}_{JC}t}$ or the influence of $g$. In particular, our algorithm only controls $(\omega_q,t)$. These control knobs are easier to adjust experimentally than the 3 degrees of freedom including two-qubit gates suggested by PGH.

A numerical challenge with Bayesian inference techniques is that each application of Bayes' theorem (\ref{eq:bayes_theorem}) requires the evaluation of the computationally expensive integral $P(d)$. To calculate the integrals we adapt a sequential Monte Carlo approach \cite{west93,gordon93,liu00,huszar12,granade12} with moving grid points or ``particles'' whose density describes the probability distribution of interest \cite{modifications}. In evaluating a probability distribution in fixed grid points, the density of the grid points would limit the precision on the estimates, which is mitigated by the adaptive grid. We perform the calculations with 50 000 particles, so, for instance, in computing $P(d)$ one evaluates 50 000 times the likelihood function.

To consider the performance of our algorithm, we have applied it to ensembles of 10 000 simulated samples with randomly chosen parameters $(g_0,\omega_{r,0})$. Here the subscript $0$ denotes a specific fixed true value, in contrast to the symbols naming a quantity. The values $g_0$ have been chosen from a log-normal distribution with the mean $\mu_g$ and the standard deviation $\sigma_g$ while $\omega_{r,0}$ have been chosen from the normal distribution $\mathcal{N}(\mu_\omega$,$\sigma_\omega$). For each sample we have chosen the initial prior of the $(g,\omega_r)$ estimate to coincide with the probability distribution from which the true values $(g_0,\omega_{r,0})$ are randomly picked. Unless specified otherwise we have chosen $\sigma_g=0.25\mu_g$ and $\sigma_\omega=\mu_g$.

Figure~\ref{fig:fig3} exhibits the relative median squared error as a function of measurement shots for $g$ (solid) and $\omega_r$ (dotted). The changes in parameters represent the effects of relaxation and noise: (i) $T_1=\infty$, $P_e=0$ (black), (ii) $T_1=\infty$, $P_e=0.1$ (blue),  (iii) $T_1=\frac{2000\pi}{\mu_g}$, $P_e=0.1$ (red), and (iv) $T_1=\frac{40\pi}{\mu_g}$, $P_e=0.1$ (brown). Here $T_1$ is the relaxation time and $P_e$ the probability of a readout error. The error decreases exponentially with the base of the exponential function affected by $P_e$. When the error reaches a low crossover value determined by $T_1$, the decay law crosses over into a power law.
\\
\begin{table} 
\begin{center} 
\caption{Number of outliers per 10 000 simulated samples with $T_1=\infty$, $P_e=0$. Rows correspond to the number of outliers with squared error larger than $\tilde{\mathcal{E}}^2_g$ after a given number of measurement shots (indicated by the columns).}
\begin{tabular}{c ||  c | c | c | c  | c | c}
$\tilde{\mathcal{E}}^2_g$/Shots & 150 & 300 & 600 &  1200 & 3600 & 6600 \\ 
\hline\hline
$10^{-10}$ & 533 & 466 &  276 & 25 & 3 & 0\\
\hline
$10^{-7}$ & 265 & 251 & 111 & 18 & 2 & 0\\
\hline
$10^{-4}$ & 118 & 116 & 25 & 14 & 1 & 0\\
\hline
\end{tabular} 
\label{table1} 
\end{center} 
\end{table} 
For each simulated ensemble there are some samples we call ``outliers'' for which the error significantly exceeds
the exponentially decaying median and the width of the posterior distribution. The outliers mostly correspond to unlikely parameter values. Our algorithm can be made robust against such outliers through repetition as follows, based on the idea that most bad estimates are good initial guesses. After 300 measurement shots we set the prior widths back to their original values but keep the mean of the probability distribution unchanged. Another 300 measurement shots are performed thereafter. We then compare the estimates after 300 and 600 measurement shots. If their difference is smaller than a set threshold, we conclude that we have found a correct estimate, otherwise we start a new search of the estimate. For the new search we choose a prior whose mean values for $g$ and $\omega_r$ are randomly chosen from the original prior while the prior widths equal those of the original prior. Table I summarizes the performance of our outlier correction scheme. Outliers are defined as the samples with the squared error of the $g$ estimate larger than threshold $\tilde{\mathcal{E}}_g^2$. Our scheme appears to reduce the number of outliers with an acceptable overhead.

Figure~\ref{fig:fig4} exhibits the average number of measurement shots required to meet the desired level of relative mean squared error. Each point represents the average performance over 10 000 samples. The robustness of our algorithm against measurement errors and relaxation is demonstrated in Figs.~\ref{fig:fig4}(a) and~\ref{fig:fig4}(b), and the effect of greater initial parameter uncertainty is considered in Figs.~\ref{fig:fig4}(c) and~\ref{fig:fig4}(d). We find that when $|\omega_{r,0}-\mu_\omega|$ is no larger than few times $g_0$, the number of measurement shots is comparable to the one ideally required when $\omega_r$ is initially known precisely \cite{ferrie13,granade12}, indicating the near optimality of our algorithm in this parameter region.

With conventional swap spectroscopy, the choice of the measurement settings tends to limit the accuracy of the estimates. Their squared error scales no better than $\sim M_s^{-1}$ in the number of measurement settings 
$M_s$ (grid points distributed on a rectangular grid in Fig.~\ref{fig:fig1}) and in a typical experiment $M_s$ is in the range  $O(10^4)- O(10^5)$. The accuracy of the estimate of $P_{t,\omega_q}(1|g,\omega_r)$ [color scale in Fig.~\ref{fig:fig1}(a)] is also limited due to projection noise as the ensemble average is typically taken over $M_{r}\sim O(10^3)- O(10^4)$ trials. Our results show that it is possible to obtain far more accurate estimates with a much smaller number of measurement shots which makes our approach highly efficient.

In order to estimate the experimental time advantage, one has to take into account the delay of requesting and transferring data between the experiment and the computation that determines the measurement setting $t_{\rm latency}$, as well as the time to computationally determine a single measurement setting $t_{\rm adapt}$. If the adaptive scheme is implemented on a field-programmable gate array, $t_{\rm latency}\sim$ 200 $\mu$s is dominated by $t_{\rm adapt}\sim 10$ ms \cite{private_communication}. The typical time to initialize the system and perform the measurement, e.g., in superconducting qubits is $t_{\rm cycle}\sim$ 100 $\mu$s. We therefore expect it is optimal to repeat measurements in a single setting $\sim 10-100$ times. Making several repetitions decreases the number of required adaptation steps. We estimate that in a few seconds it is possible to obtain estimates that are more precise by several orders of magnitude compared to conventional swap spectroscopy where the time to carry out $M_s M_{r}$ measurement shots ranges from minutes to several hours. For experimental considerations it will also be useful to develop a more general algorithm to choose measurement settings for wide frequency ranges  containing multiple resonators at frequencies $\omega_r^{(i)}$. 

In conclusion, by adaptively focusing measurements on the regions of high information gain we obtained a globally efficient measurement strategy for two control parameters. Our algorithm makes advanced spectroscopy drastically more efficient.

We thank D. Sank, J. Kelly, and J.M. Martinis for numerous discussions. F.W. acknowledges early discussions with C. Bender, H. Z\"ahle, and M. Hein. Work supported by the European Union through ScaleQIT as well as by the Office of the Director of National Intelligence (ODNI), Intelligence Advanced Research Projects Activity (IARPA), through the Army Research Office. All statements of fact, opinion or conclusions contained herein are those of the authors and should not be construed as representing the official views or policies of IARPA, the ODNI, or the U. S. Government.

\pagebreak
{\large\bf Supplemental Material}\\

The dynamics of the qubit-resonator system is governed by the Master equation
\begin{eqnarray}
\dot{\hat{\rho}}_S=-i\left[\hat{H},\hat{\rho}_S\right]+\mathcal{D}\left[\hat{A}\right]\hat{\rho}_S,
\end{eqnarray}
with ${\hat{\rho}}_S$ the density matrix in the Schr\"odinger picture.
In the absence of relaxation the system is described by the Jaynes-Cummings Hamiltonian
\begin{eqnarray}
&\hat{H}=\hat{H}_0+\hat{H}_1,\nonumber\\
&\hat{H}_0=\frac{E_{q}}{2}\hat{\sigma}_{z}+\frac{E_{r}}{2}\hat{\tau}_{z},\quad \hat{H}_1 = g\left(\hat{\sigma}_{+}\hat{\tau}_{-}+\hat{\sigma}_{-}\hat{\tau}_{+}\right)\label{eq:Hamiltonian}.
\end{eqnarray}
Here $E_q$ and $E_r$, are the bare qubit and resonator energies, respectively, and the qubit-resonator coupling
is characterized by $g$. Furthermore, $\hat{\sigma}_{z}$ $(\hat{\tau}_{z})$ are Pauli matrices in qubit (resonator) subspace, whereas
$\hat{\sigma}_{\pm}$ and $\hat{\tau}_{\pm}$ are raising and lowering operators.
We focus on energy relaxation of the qubit as the main channel of
loss, described by a Lindblad form 
\begin{equation}
\mathcal{D}\left[\text{\ensuremath{\hat{A}}}\right]\hat{\rho}_S\equiv\hat{A}\hat{\rho}\hat{A}^{\dagger}-\frac{1}{2}\left\{ \hat{A}^{\dagger}\hat{A},\hat{\rho}_S\right\} \label{eq:lindblad_form}
\end{equation}
with the amplitude damping Lindblad operator
\begin{equation}
\hat{A}=\sqrt{\Gamma}\hat{\sigma}_{-}\label{eq:relaxation_operator}
\end{equation}
where $\Gamma$ is the relaxation rate.

We would first like to show that we can drop the state $|11\rangle$ in
the equations from the very beginning. We switch to the interaction picture
with the Schwinger-Tomonaga equation 
\begin{equation}
\dot{\hat{\rho}}_I=-i\left[\hat{H_1},\hat{\rho}_I\right]+\mathcal{D}\left[\hat{A}_{I}\right]\hat{\rho}_I.
\end{equation}
It is easy to show that $\hat{H}_{I}|11\rangle\text{=0}$. Moreover, $\hat{A}_{I}=e^{-iE_qt}\hat{A}$ 
implies $\hat{A}_I\rho_I\hat{A}_I^{\dagger}|11\rangle=0$. Thus all the time-dependence left is
\begin{equation}
\left\langle 11\left|\dot{\hat{\rho}}_I\right|11\right\rangle =
-\Gamma\left\langle 11\left|\hat{\rho}_I\right|11\right\rangle.
\label{eq:auton}
\end{equation}
Therefore if the initial state does not include a $|11\rangle$ component, that state stays unoccupied.

Below we assume that the initial state at $t=0$ is $|10\rangle$. As the Jaynes-Cummings Hamiltonian preserves the number of excitations we can describe the resulting dynamics in the single-excitation subspace
by the Hamiltonian $\hat{H}'=\frac{\Delta\omega}{2}\hat{\eta}_{z}+g\hat{\eta}_{x}$.
Here $\hat{\eta}$ are Pauli matrices and $\Delta\omega=\omega_{q}-\omega_{r}$ is the detuning frequency.
The eigenstates of  $\hat{H}'$ are
\begin{eqnarray}
&|e\rangle  =  \cos\frac{\theta}{2}|10\rangle+\text{\ensuremath{\sin}}\frac{\theta}{2}|01\rangle,\nonumber\\
&|g\rangle =  -\sin\frac{\theta}{2}|10\rangle+\cos\frac{\theta}{2}|01\rangle,
\end{eqnarray}
where $\tan\theta=\frac{2g}{\Delta\omega}$. The eigenenergies are $E=\pm\frac{\hbar\omega_R}{2}$, with
$\omega_R = \sqrt{\omega^{2}+4g^{2}}$. Note that $\Delta\omega=0$ corresponds to
$\theta=\pi/2$. 

To treat relaxation, we now express the qubit lowering operator
in these eigenstates as 
\begin{eqnarray}
\begin{split}
\hat{\sigma}_{-} =& |0\rangle\langle1|\otimes\hat{1}\\
 =&|00\rangle\left(\cos\frac{\theta}{2}\langle
   e|-\sin\frac{\theta}{2}\langle g|\right)+\\
&+\left(\sin\frac{\theta}{2}|e\rangle+\cos\frac{\theta}{2}|g\rangle\right)\langle11|.
 \end{split}
\end{eqnarray}
Our approach is to perform a rotating wave approximation (RWA),
recognizing that $\omega_R/2\ge g\gg\Gamma$. 
To carry out the transformation, we first introduce a further interaction picture where the perturbation now includes only the relaxation
\begin{equation}
\dot{\hat{\rho}}_{I'}=\mathcal{D}\left[\hat{A}_{I'}\right]\hat{\rho}_{I'}.
\label{eq:new_int_pict}
\end{equation}
In this picture the lowering operator in the qubit subspace is
\begin{eqnarray}
\begin{split}
\hat{\sigma}_{-}^{(I')}=&e^{-i(E_{q}+E_r)t/2\hbar}|00\rangle\left(\cos\frac{\theta}{2}e^{-i\omega_Rt/2}\langle e|\right.\\
&\left.-\sin\frac{\theta}{2}e^{i\omega_Rt/2}\langle g|\right)+e^{-i(E_{q}+E_r)t/2\hbar}\\
&\times\left(\sin\frac{\theta}{2}e^{i\omega_Rt/2}|e\rangle+\cos\frac{\theta}{2}e^{-i\omega_Rt/2}|g\rangle\right)\langle11|.
\end{split}
\end{eqnarray}

To evaluate the Lindblad form on the right hand side of Eq.~\eqref{eq:new_int_pict} we need to transform the operator $\hat{\sigma}_{+}\hat{\sigma}_{-}=|1\rangle\langle1|\otimes\hat{1}$ to the interaction picture.
In the RWA the transformed operator reads 
\begin{equation}
\left(\hat{\sigma}_{+}^{(I')}\hat{\sigma}_{-}^{(I')}\right)_{{\rm RWA}}=\cos^{2}\frac{\theta}{2}|e\rangle\langle e|+\sin^{2}\frac{\theta}{2}|g\rangle\langle g|+|11\rangle\langle11|.
\end{equation}

We can now drop all the terms involving the state $|11\rangle$ using the same argument as in the context of 
Eq.~\eqref{eq:auton}. Hence we obtain
\begin{eqnarray}
\left\langle e\left|\dot{\hat{\rho}}_{I'}\right|e\right\rangle  & = & -\cos^{2}\frac{\theta}{2}\Gamma\left\langle e\left|\hat{\rho}_{I'}\right|e\right\rangle,\nonumber \\
\left\langle g\left|\dot{\hat{\rho}}_{I'}\right|g\right\rangle  & = & -\sin^{2}\frac{\theta}{2}\Gamma\left\langle g\left|\hat{\rho}_{I'}\right|g\right\rangle,\nonumber \\
\left\langle e\left|\dot{\hat{\rho}}_{I'}\right|g\right\rangle  & = & -\frac{\Gamma}{2}\left\langle e\left|\hat{\rho}_{I'}\right|g\right\rangle.
\label{eq:rates}
\end{eqnarray}

We would now like to calculate the evolution of the population of $|10\rangle$ starting in that state. The initial  density matrix is 
\begin{eqnarray}
&&\hat{\rho}_{I'}(t=0)=|10\rangle\langle10|\nonumber\\
&&=\cos^{2}\frac{\theta}{2}|e\rangle\langle e|+\sin^{2}\frac{\theta}{2}|g\rangle\langle g|-\frac{1}{2}\sin\theta\left(|e\rangle\langle g|+|g\rangle\langle e|\right).\nonumber\\
\end{eqnarray}
Using Eqs.~\eqref{eq:new_int_pict} and \eqref{eq:rates} we find that the state decays according to
\begin{eqnarray}
\begin{split}
\hat{\rho}^{I}(t)=&\cos^{2}\frac{\theta}{2}e^{-\Gamma t\cos^{2}\theta/2}|e\rangle\langle e|+\sin^{2}\frac{\theta}{2}e^{-\Gamma t\sin^{2}\theta/2}|g\rangle\langle g|\\
& -\frac{1}{2}\sin\theta e^{-\Gamma t/2}\left(|e\rangle\langle g|+|g\rangle\langle e|\right).
\end{split}
\end{eqnarray}

The projection operator $\hat{\Pi}=|10\rangle\langle10|$ corresponding to the measurement
is in the interaction picture 
\begin{eqnarray}
\begin{split}
\hat{\Pi}_{I'}(t)=&\cos^{2}\frac{\theta}{2}|e\rangle\langle e|+\sin^{2}\frac{\theta}{2}|g\rangle\langle g|\\
&-\frac{1}{2}\sin\theta\left(|e\rangle\langle g|e^{i\omega_Rt}+|g\rangle\langle e|e^{-i\omega_Rt}\right).
\end{split}
\end{eqnarray}

Since the trace of an operator is the same in the interaction and the Schr\"odinger pictures we find the occupation probability of the state $|10\rangle$
\begin{eqnarray}
\begin{split}
&P_{t,\omega_q}\left(1|g,\omega_{r}\right)={\rm Tr}\left(\hat{\Pi}\hat{\rho}\right)=\frac{\left(1+\cos\theta\right)^{2}}{4}
e^{-\Gamma t\left(1+\cos\theta\right)/2}
\\
&+\frac{(1-\cos\theta)^{2}}{4}
e^{-\Gamma t(1-\cos\theta)/2}+\frac{\sin^{2}\theta}{2}e^{-\Gamma t/2}\cos \omega_Rt.
\end{split}
\end{eqnarray}
Using $\tan\theta=\frac{2g}{\Delta\omega}$ and $T_1=\frac{1}{\Gamma}$ this can be recast as
\begin{eqnarray}
\begin{split}
&P_{t,\omega_q}\left(1|g,\omega_{r}\right)=\left(\frac{\omega_R+\Delta\omega}{2\omega_R}\right)^{2}e^{-(\omega_R+\Delta\omega)t/2\omega_RT_1}\\&+\left(\frac{\omega_R-\Delta\omega}{2E}\right)^{2}e^{-(\omega_R-\Delta\omega)t/2\omega_RT_1}+\frac{2g^{2}}{\omega_R^{2}}e^{-t/2T_1}\cos \omega_Rt.
\end{split}
\end{eqnarray}
This generalizes Eq.~(2) in the main text for finite relaxation.

\end{document}